\documentclass[aps,prb,twocolumn,amsmath,amssymb,superscriptaddress,showpacs]{revtex4-1}

\newcommand{\be}{\begin{equation}}
\newcommand{\ee}{\end{equation}}
\newcommand{\bea}{\begin{eqnarray}}
\newcommand{\eea}{\end{eqnarray}}

\renewcommand{\vec}{\mathbf}

\renewcommand{\vr}{\vec{r}}

\newcommand{\vsigma}{\mbox{\boldmath $\sigma$}}

\newcommand{\vA}{\vec{A}}
\newcommand{\vB}{\vec{B}}
\newcommand{\vnabla}{\mbox{\boldmath $\nabla$}}

\newcommand{\vp}{\vec{p}}

\newcommand{\vF}{v_{\rm F}}

\usepackage[utf8]{inputenc}
\usepackage{amsfonts}
\usepackage{amsthm}
\usepackage{fontenc}
\usepackage{graphicx}
\usepackage{textcomp}
\usepackage{epstopdf}
\usepackage{braket}
\usepackage{mathtools}
\usepackage{dcolumn}
\usepackage{bm}
\usepackage{epsfig}
\usepackage[usenames,dvipsnames]{color}
\usepackage{booktabs}
\usepackage{lipsum}
\usepackage[colorlinks=true,citecolor=blue,linkcolor=blue]{hyperref}
\bibpunct{[}{]}{,}{n}{}{}

\begin{document}
\newcommand{\abs}[1]{\lvert#1\rvert}
\title{Valley polarization braiding in strained graphene}
\author{D. Faria}
\affiliation{Instituto Polit\'{e}cnico, Universidade do Estado do Rio de Janeiro, Nova Friburgo, Rio de Janeiro 28625-570, Brazil}
\affiliation{Department of Physics and Astronomy, Ohio University, Athens, Ohio 45701-2979, USA}
\author{C. Le\'{o}n}
\affiliation{Instituto de F\'{i}sica, Universidade Federal Fluminense, Niter\'{o}i,  Rio de Janeiro 24210-340, Brazil}
\author{L. R. F. Lima }
\affiliation{Instituto de F\'{i}sica, Universidade Federal Fluminense, Niter\'{o}i,  Rio de Janeiro 24210-340, Brazil}
\affiliation{
Departamento de F\'{i}sica, Universidade Federal Rural do Rio de Janeiro, Serop\'{e}dica, Rio de Janeiro 23897-000, Brazil}
\author{A. Latg\'{e}}
\affiliation{Instituto de F\'{i}sica, Universidade Federal Fluminense, Niter\'{o}i,  Rio de Janeiro 24210-340, Brazil}
\author{N. Sandler}
\affiliation{Department of Physics and Astronomy, Ohio University, Athens, Ohio 45701-2979, USA}

\date{\today}

\begin{abstract}

Previous works on deformed graphene predict the existence of valley-polarized states, however, optimal conditions for their detection remain challenging. We show that in the quantum Hall regime, edge-like states in strained regions can be isolated in energy within Landau gaps.  We identify precise conditions for new conducting edges-like states to be valley polarized, with the flexibility of positioning them at chosen locations in the system.  A map of local density of states as a function of energy and position reveals a unique braid pattern that serves as a fingerprint to identify valley polarization. 

\end{abstract}


\maketitle 

Strained graphene has emerged as an important tool to implement valleytronic based devices, and in particular, in protocols for quantum computation\cite{Zhang2017, Ghaemi2013, Rainis2011, Ramon2016, Dawei2018, Settnes2016, triaxialroy, Settnes2017, Settnes2017, Shubnyi2018, Vanessa2018, Bahamon2019, Ang2017}. Recent experimental developments show that substrate engineering can be used to design deformation geometries with specific strain profiles\cite{Scott2008,Lindahl2012,Stroscio2012,Mason2018,Andrei2017,Andrei2019,Goldsche2018,Alexander2017,Li2015,Li2019,Jia2019}. Clear signatures of valley splitting in confined geometries represent an important step in this direction, as exemplified by STM studies on graphene quantum dots \cite{Nils2016}. In more extended configurations, similar observations have been reported on multiple fold structures\cite{Li2015,Li2019} with preliminary evidence of valley polarized states. These studies are supported by previous work on extended deformations predicting valley polarized edge-like states at the strain region, which acts as a waveguide focusing electron currents\cite{Zhang2017, Ghaemi2013, Rainis2011, Ramon2016, Dawei2018}. These are all promising structures for potential device applications. However, several drawbacks are still present because optimal conditions for creation and detection of valley split currents are not well-defined. 

To take advantage of the existence of valley polarized channels, usually embedded in graphene's conducting background, it is crucial to separate their contribution from other extended states. We show that this can be achieved by introducing an external magnetic field large enough to take the system into the Quantum Hall regime. Such a configuration conveniently allows the isolation of the valley polarized edge states in energy and in real space. 
As we show below, it is possible to design configurations within available experimental capabilities to produce valley polarized currents for a wide energy range within Landau gaps. Moreover, the flexibility to place the deformation at different parts of the sample provides a wider versatility of contact probes to identify and collect these currents. 

We present local density of states (LDOS) results for a model of graphene with a fold-like deformation that predict valley split peaks that could be measured in STM experiments. As the deformed region is traversed across, maximum LDOS intensities for each valley evolve in energy, leading to a braid structure that serves as a unique fingerprint of valley polarized states. Under bias, these states generate new extra conducting channels that can be visualized as new edge states created along the deformation region.
  
In order to bring attention to the interplay between deformation parameters and magnetic length, we perform combined analytical and numerical studies based on the continuum and tight-binding descriptions of electrons in graphene. As we are interested in the Quantum Hall regime, the deformation is considered as a perturbation to Landau Level states. Our results show the existence of two distinct regimes characterized by $\gamma = l_B/b$, i.e. the ratio between the magnetic length $\l_B$ and the deformation width $b$. For $\gamma >1$ the broad Landau level states average over the deformed region. In contrast, for $ \gamma < 1$ the magnetic confinement allows the electrons to follow the inhomogeneous profile introduced by strain. In this last regime, the spatial separation between the polarized currents becomes larger. This could encourage the design of devices where contacts can efficiently detect  polarized currents with a potential use as logic gates in quantum computing devices \cite{Ang2017}.

\noindent{\it Model}  The electronic properties of strained graphene in the presence of a magnetic field are described by the nearest neighbor tight-binding Hamiltonian\cite{Castro2009} 
\begin{equation}
H=\sum\limits_{<i,j>} t_{ij}c_i^\dagger c_j + h.c.\,\,,
\label{eq:HTB}
\end{equation}
where $c_i^\dagger$ ($c_i$) is the creation (annihilation) field operator in the $i$-th site. The modified nearest-neighbor hopping energy, $t_{ij}$, is  given by\cite{Castro2009,Goerbig}
\begin{equation}
t_{ij} = t_{0}e^{i\Delta\phi_{i,j}}e^{-\beta \left(\frac{l_{ij}}{a_{cc}}-1\right)} \,\,\, ,
\end{equation}
with $\beta=|\partial \log t_0/\partial \log a_{cc}|\approx 3$,  and $t_0$ and $a_{cc}$ are the hopping parameter and the lattice constant of pristine graphene.
The magnetic field is included via the Peierls substitution, $\Delta\phi_{i,j}=2\pi(e/h)\int_{\vr_j}^{\vr_i}\vA\cdot d\vr$, with $\vr_i$ and $\vr_j$ nearest neighbors positions. The strain field, given in terms of the elasticity tensor $\varepsilon$,  modifies interatomic distances $l_{ij}=\frac{1}{a_{cc}}\left(a^{2}_{cc}+\varepsilon_{xx}x_{ij}^2+\varepsilon_{yy}y_{ij}^2+2\varepsilon_{xy}x_{ij}y_{ij}\right)$, where $x_{ij}$ and $y_{ij}$ correspond to the projected distance between sites $i$ and $j$ before the deformation. In these expressions the $x$-axis is chosen along the zigzag direction. 
At low energies the effective continuum Hamiltonian is given by two copies of a 2D Dirac equation $H^{D}_{K(K')}=\vF \vsigma \cdot \vp $ written in the valley symmetric representation. Here
$\vF\approx 10^6 m/s$ is the Fermi velocity\cite{defFermivelocity},  $\vsigma=(\sigma_x,\sigma_y)$ are Pauli matrices acting on the pseudospin degree of freedom associated with the sublattice $(A, B)$ structure of the honeycomb lattice~\cite{Castro2009}, and $\vp$ the electronic momentum around the K (K') point. The magnetic field is implemented using the minimal coupling $\vp=\vp+e\vA$ in the Landau gauge, as $\vA=B(-y,0)$. 
The unstrained system has relativistic Landau levels (LLs)  given by $E_N=\pm\frac{\hbar v_{F}}{l_{B}}\sqrt{2N}$ with the $\pm$ representing conduction and valence bands, respectively. The magnetic length is given by $l_B= \sqrt{\frac{\hbar}{eB}}$, and $N$ is the integer label for each Landau level. 

\begin{figure*}
\includegraphics[scale=0.70]{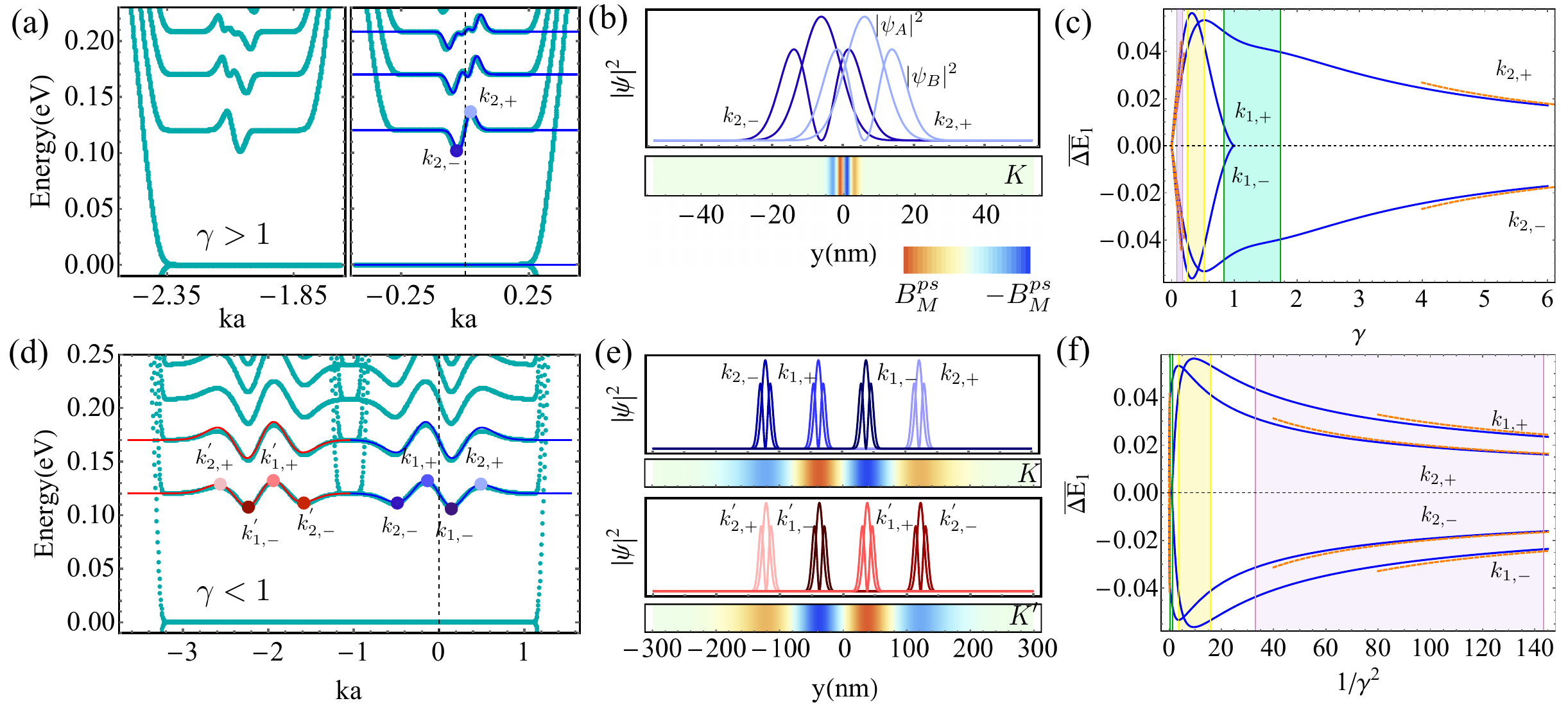}
\caption{(Color online)  (a) and (d) Comparison between continuum (blue (K) and red (K')) and tight-binding (cyan) band structure results for deformed graphene in the QH regime. (b) and (e) Probability density distributions for $k$ states identified in (a), and (d).  Bottom: contour plot of pseudomagnetic field with maximum values of $B^{ps}_{M} =82T$ (b), and $B^{ps}_{M} =2.3T$ (e). Parameters: (a) and (b) $\gamma=2.7$, $b=20a_{cc}$, $\xi = 0.2$, maximum strain $\varepsilon_\text{M}=1.5\%$  and (d) and (e) $\gamma=0.07$, $b=800a_{cc}$, $\xi = 0.22$, $\varepsilon_\text{M}=1.7\%$. $B=11$T for both cases. Panels (c) and (f): Scaling of the first LL energy correction  ${\Delta E}_{1}$ in units of $(\xi^2 t')$,  as a function of $\gamma$ and $1/\gamma^2\propto B$, respectively. Continuum (blue) lines represent exact evaluation of energy corrections while dashed (orange) lines correspond to analytic expressions in the asymptotic regimes given by Eq.~(\ref{eq:deltaE1}).} 
\label{Fig1}
\end{figure*}

To study the effects of strain in this regime we chose to represent a model for a non-uniform strain, introduced by a fold-like deformation with a height-profile $h(\vr)$ written generically as 
\be
h\left( y\right) =h_0
e^{-\frac{(y-y_{0})^2}{b^2}},
\label{gaussian}
\ee
where $h_0$ and $b$ describe amplitude and effective extension of the fold, respectively, and $y_{0}$ indicates the position of its center. In the continuum limit, the corresponding strain tensor  $\epsilon_{ij}=\frac{1}{2} \partial_i h \partial_jh $ gives rise to the  pseudo gauge field\cite{Ando} 
\begin{equation}
\left(\begin{array}{c}
A^{ps}_{x}\\
A^{ps}_{y}
\end{array}\right)=\left(\begin{array}{c}
\varepsilon_{xx}-\varepsilon_{yy}\\
-2\varepsilon_{xy}
\end{array}\right)= \left(\begin{array}{c}
-2\frac{y^{2}}{b^{4}}h(y)^{2}\\
0
\end{array}\right) \,\,\, ,
\end{equation}
and a pseudo magnetic field $\vB^{ps}_{K(K')}= \pm\frac{\Phi_0}{(2\pi)} \left(\frac{-\beta}{2a_{cc}}\right) \vnabla \times \vA^{ps}$, with $+ (-)$ for valley K (K'), where $\Phi_0$ is the unit of quantum flux. The model, chosen to emphasize the spatial dependence of $\vB^{ps}$ shown in the contour plots at the bottom of Fig.~\ref{Fig1} (b) and (e), reveals physical features  that are determined by the extension of the deformed region $b$. These features should be observable in samples with more general non-uniform strain profiles, thus making our predictions relevant for a broad range of experimental setups.

The electron dynamics is governed by: 
\begin{equation}
H_{K(K')}=\hbar \vF \vsigma \cdot \left(-i\vnabla -\frac{e}{\hbar}\vA \pm \frac{\beta}{2a_{cc}} \vA^{ps}\right) \,\,\, .
\label{eq:HKK'}
\end{equation}
Since we are interested in the Quantum Hall regime, the gauge field due to the deformation is treated as a perturbation. 
Because of the $x-$direction translation invariance, Eq.~(\ref{eq:HKK'}) allows solutions of the form $\Psi(x,y)= \psi(y)e^{ikx}$. The effective one-dimensional Hamiltonian reduces to
\begin{equation}
\label{Dequation}
\left( \begin{array}{cc}0 & \hat{\mathcal{O}}\mp t' (\xi \gamma)^2\varepsilon_{\tilde{y}\tilde{y}}\\ \hat{\mathcal{O}^{+}}\mp t' (\xi \gamma)^2\varepsilon_{\tilde{y}\tilde{y}}& 0\end{array}  \right)  \psi(\tilde{y})= E\psi(\tilde{y}) \,\,\, ,
\end{equation}
with $\hat{\mathcal{O}}=\frac{\hbar \omega_c}{\sqrt{2}}(\partial_{\tilde{y}}+\tilde{y})$, $\hat{\mathcal{O}^{+}}=\frac{\hbar \omega_c}{\sqrt{2}}(-\partial_{\tilde{y}}+\tilde{y})$ and $\omega_c=\frac{\sqrt{2}\vF}{l_B}$. Dimensionless coordinates are defined as $\tilde{x}=x/l_B$, $\tilde{y}=(y/l_B-\tilde{k})$, and the effective hopping $t' =\beta \hbar v_F/a_{cc} \approx 13.9eV$. The deformation parameter $\xi = \left(h_0/b\right)$ characterizes the strain intensity $\varepsilon_M = \xi^2/e$, and the dimensionless strain tensor is thus given by $\varepsilon_{\tilde{y}\tilde{y}}= (\tilde{y}+\tilde{k})^2e^{2(\tilde{y}+\tilde{k})^2\gamma^2}$, with $\tilde{k}=kl_B$. 

The analysis of the continuum model is done with perturbation theory techniques for energy and eigenstates, in terms of the composite parameter $g = \xi \gamma \ll 1$ for each value of $\tilde{k}$  and $\tilde{y}$. Since $\gamma$ can be smaller or larger than 1, the above definition implies the existence of two different regimes: $1\ll \gamma \ll 1/\xi$ and $\gamma \ll 1/\xi \ll 1$, both tractable within perturbation theory.
The unperturbed spinor eigenstates are given by
 \begin{equation}
\label{Psi0}
\Psi^0(\tilde{x},\tilde{y})=\frac{1}{\sqrt{2}}\sqrt{\frac{l_B}{L_x}}\left(\begin{array}{c}
\psi_{N-1}(\tilde{y})\\
\pm \psi_{N}(\tilde{y})
\end{array}\right)e^{i\tilde{k}\tilde{x}} \,\,\, ,
\end{equation}
where $\psi_{N}(\tilde{y})=2^{N/2}N!e^{-\tilde{y}^2 / 2}H_N(\tilde{y})$ , $H_N(\tilde{y})$ is the Hermite polynomial of N-th order, $L_x $ ($\to \pm \infty$) rises from the normalization of the plane wave and $\pm$ corresponds to positive and negative energies, respectively.

The change in the energy of the $N$-th Landau level, $\Delta E_N$, is given by 
\begin{equation}
\Delta E_N(\tilde{k}) =-t' (\xi \gamma)^2\int_{-\infty}^{\infty}\varepsilon_{\tilde{y}\tilde{y}}( \tilde{y},\tilde{k}) \,\psi_{N-1}(\tilde{y})\,\psi_{N}(\tilde{y})d\tilde{y} \,\,\, .
\label{eq:deltaEN}
\end{equation}
The analytic solution of the integral provides an exact expression for the energy corrections and  asymptotic expressions for the reduced gaps can be derived (see Fig.~\ref{Fig1} and Supp. Mat.\cite{SM}). For a fixed  strain value ($\xi =$ const.), the first Landau level corrections for $\gamma \ll 1$ and  $\gamma \gg 1$ are:
\be \Bigg\{
\begin{array}{ll}
\Delta E_1 = \pm c_1 t' \xi^2\gamma^{-1} + O[\gamma^{-3}] , \,\,\, \text{for}  \,\,\, \gamma \gg 1 \,\,\, ,
\\ \Delta E_1 = \pm c_{2(3)} t' \xi^2 \gamma + O[\gamma^{3}],  \,\,\, \text{for}  \,\,\, \gamma \ll 1 \,\,\,.
\label{eq:deltaE1}
\end{array}
\ee
with constant values  $c_1\approx 0.1$, $c_2\approx  0.2$ and $c_3\approx 0.3$.  

These expressions are consistent with numerical results obtained by solving (\ref{eq:HTB}) for a nanoribbon geometry with zigzag termination along the fold axis direction. The ribbon widths were chosen to avoid edges effects. Although valley polarization is obtained  for systems with deformations placed off-center or asymmetric profiles\cite{triaxialroy,Li2019,Settnes2016,Settnes2017}, we show that valley polarized currents exist even in perfect symmetric configurations in the appropriate regimes. 

\noindent{\it Results.} Fig.~\ref{Fig1} shows a comparison between continuum and tight-binding for a fixed external magnetic field $B= 11T$. Panels (a) and (d) show band structure results for the regimes $\gamma > 1$ and $\gamma < 1$ respectively, with parameters appropriate for currently available experimental realizations\cite{Lim2015,Andrei2017,Li2015,Li2019}. The general profile for both band structures shows modifications in gaps between the various LLs. As expected, the pseudo field preserves electron-hole symmetry\cite{Amorim2016,Ramon2014,Vanessa2017} and the zeroth-LL is not affected. For a given level, the two regimes  exhibit different number   of local energy minima and maxima, indicated by $(k_{1, \pm}; k_{2, \pm})$ in the first LL for the $K$ valley (results for valley $K'$ are obtained by spatial inversion). The finer structure that develops at higher LLs is produced by the inhomogeneous nature of the strain field and reveals a larger number of states being affected at higher energies. Notice the excellent agreement between analytic (blue solid line) and numerical (cyan dots) results in both regimes. In panels (b) and (e), probability densities are presented for the states color-coded by the dots in (a) and (d), on top of corresponding pseudo magnetic field contours. For $\gamma >1$, as the confinement introduced by the external field is dominant (see horizontal scale), the electronic density is spread beyond the region of the pseudo field while for $\gamma <1$ states are localized at four distinct regions following the pseudo field profile. These features are a manifestation of valley polarization in space.  

Panels (c) and (f) depict the different scaling of maxima and minima energy corrections for the first LL, $\overline{\Delta E}_{1} = \Delta E_1/(\xi^2 t')$, obtained with Eq.~(\ref{eq:deltaEN}) as function of $\gamma$ and $1/\gamma^2$ respectively (blue online). Data is presented for valley $K$ (identical results for valley $K'$). The four energy corrections for states ($k_{1, \pm}, k_{2, \pm}$) identified in panel (d) are plotted. The dependence with $\gamma$ in panel (c) shows the vanishing of the correction at $\gamma =1$ for states labeled by $k_{1, \pm}$, signaling the change in regimes from $\gamma <1$ to $\gamma > 1$. For $\gamma \gg 1$ the correction vanishes as expected because the pseudo field is concentrated in a narrower region compared to the LL confinement, even when its bigger than $B$ in magnitude. The dependence for $\gamma < 1$ is better appreciated in panel (f), where $\overline{\Delta E}_{1}$ is plotted as a function of $1/\gamma^2$. Notice that the asymptotic behavior indicates vanishing of the corrections as the pseudo field decreases in magnitude while occupying a larger region of the sample. The spreading of the pseudo field in a larger area allows for a definite resolution of its sign alternation, leading to the spatial separation of the four states. The exact solution for all values of $\gamma$ is compared with the analytic expression (Eq.~(\ref{eq:deltaE1})), shown with dashed lines (orange online), exhibiting excellent agreement in the two regimes. For $\gamma \ll 1$,  the  expression for LL energy $E_N +\Delta E_N \propto \sqrt{(B\pm B^{ps}_{M})}$, reproduces the expected scaling for an effective magnetic field smaller than $B$ \cite{SM}. 
Colored areas in Fig.~\ref{Fig1} (c) and (f) depict the transition between $\gamma \gg1$ and $\gamma \ll1$ regimes.  Notice that this transition regime can be experimentally achieved at available magnetic fields for appropriate deformation extensions.  

\begin{figure}
\includegraphics[scale=0.43]{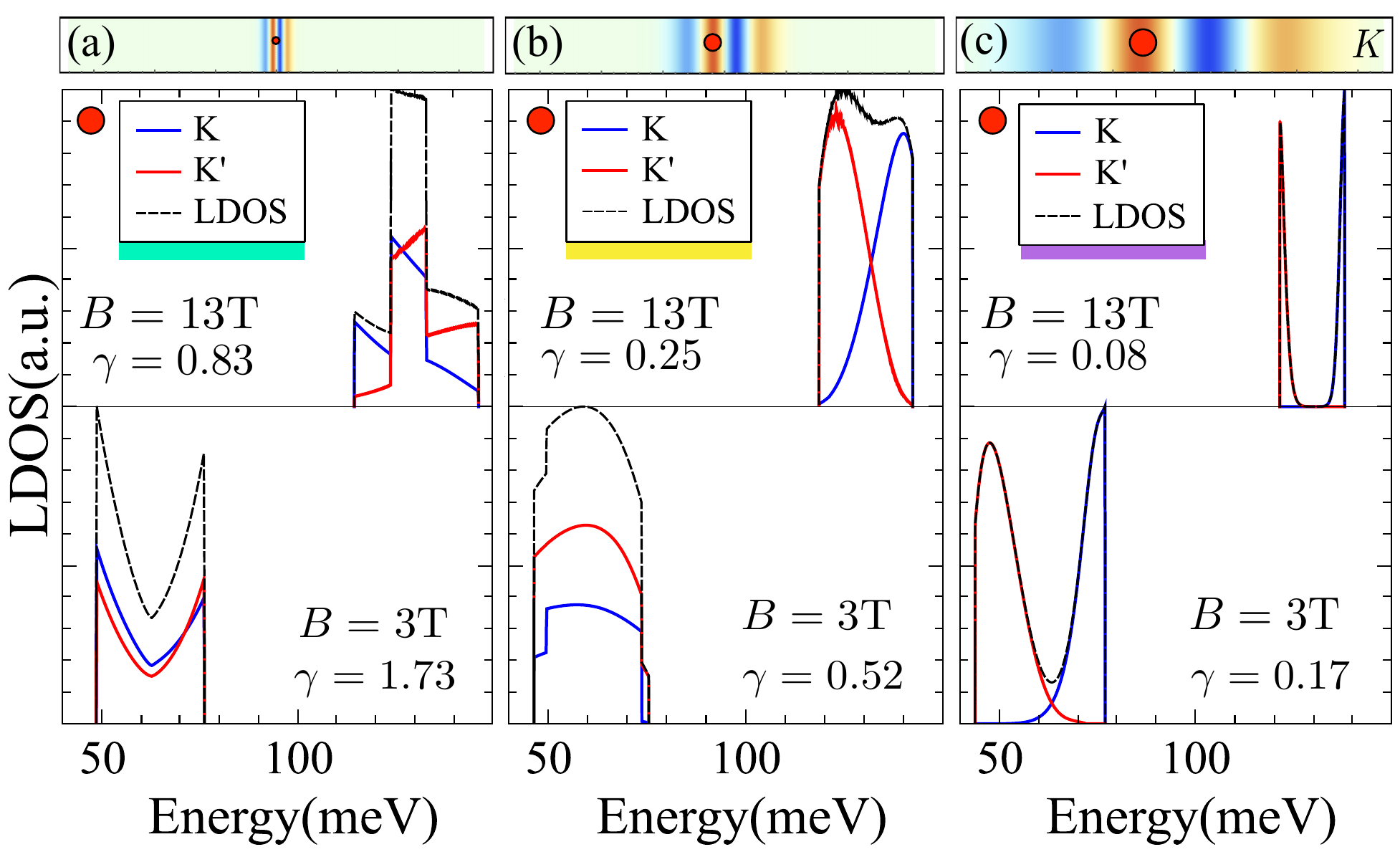}
\caption{(Color online) LDOS for K (blue) and K' (red) valleys for three deformations $(h_0,b)$:  (a) $(9 a_{cc},60 a_{cc})$, (b)  $(25 a_{cc},200 a_{cc})$, and (c) $(95 a_{cc},600 a_{cc})$, corresponding to different values of maximum strain smaller than $1\%$, and  external magnetic fields, $B = 13T$ (top panels) and $3T$ (bottom).  The pseudomagnetic-field contour plots are shown on top of the corresponding LDOS. Results obtained for positions where the pseudo field for K valley is maximum, as depicted with the red circles.}
\label{Fig2}
\end{figure}

Next, we calculate LDOS to second order in perturbation theory to provide signatures of the transition that could be observed in standard STM measurements.
Fig.~\ref{Fig2} shows results for the LDOS  for K and K' (blue and red online) valleys, for values of $\gamma$  at both boundaries of each colored shaded area in Fig.~\ref{Fig1} (c) and (f), corresponding to external fields $13T$ and $3 T$.  The LDOS is plotted at the position marked by the red dot on the pseudo magnetic field contour plots shown above, better seen in panel (c). The contours are presented for a fixed length, $2400 a_{cc}$, to emphasize the different widths $b$ used (different  $\gamma$ values). Panel (a) shows a broadened LL peak for $B = 13T$ ($\gamma = 0.83$) and a split peak for $B = 3T$ ($\gamma = 1.73$), not valley polarized. In contrast, panel (b) shows broadened peaks for both fields ($\gamma = 0.52$ and $\gamma = 0.25$). It is only for values of $\gamma \ll 1$, as shown in panel (c) ($\gamma = 0.17$ for $B=3T$ and $\gamma = 0.08$ for $B=13T$), that valley polarization is clearly resolved for both magnetic fields. Notice that the valley polarized peaks resemble van-Hove singularities representing new edge states emerging at the deformation region. In all cases, peak energies can be obtained from Eq. (\ref{eq:deltaEN}), and for a given $\gamma$ the corresponding splittings could be engineered by appropriate choice of the strain intensity.

\begin{figure} [hbt!]
\includegraphics[scale=0.46]{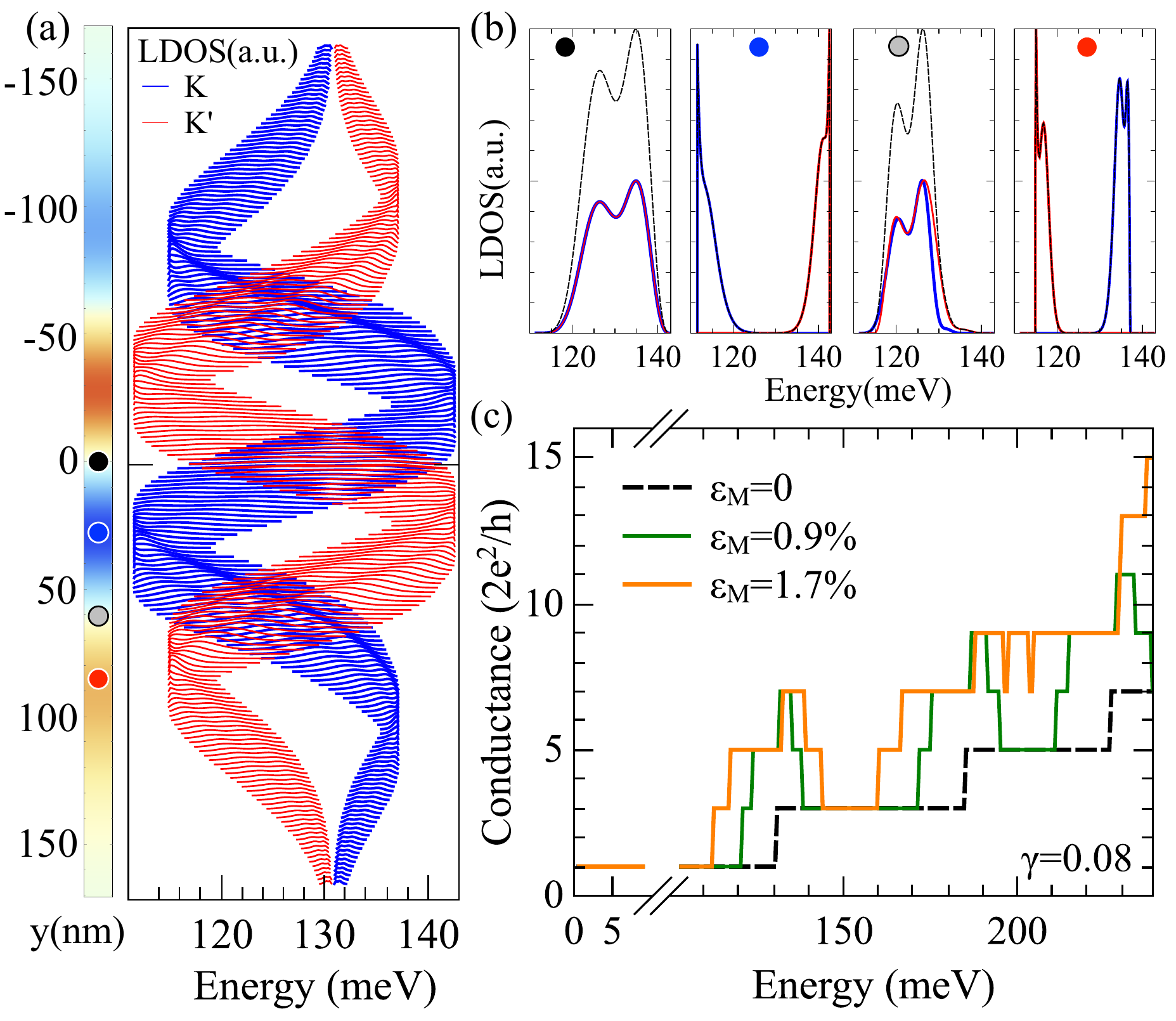} 
\caption{(Color online)  (a) LDOS as a function of energy at different position across deformation (y-direction) for K (blue) and K' (red) valleys. Curves were shifted vertically for different positions. The pseudo field profile for valley K is displayed by the colored bar.   (b)  LDOS as a function of energy for specific positions marked by colored dots in pseudo field profile. Curves were enlarged from (a) for clarity. Parameters: B=13T, deformation width $b=600 a_{cc}$, $\varepsilon_M = 1.7\%$, and $\gamma=0.08$. (c) Two terminal conductance along the deformation for a zigzag nanoribbon with B=13T.\label{Fig3}}
\end{figure}

To further investigate the dependence of valley splittings with energy, Fig.\ \ref{Fig3}(a) shows LDOS curves for different positions across the deformed region for $\gamma = 0.08$. As one moves from one side of the deformation to the other, the maxima LDOS intensities braid in a precise pattern that distinguish each valley contribution at a given spatial position.
The peak separations follow the pseudo-magnetic field profile as shown by the increased splittings around the central region. At crossings ($y = 0$ and $y = \pm 60nm$) valley polarization is strictly lost as the pseudo field vanishes at these points and valleys $K$ and $K'$ exchange places along the braid. Panel (b) highlights this evolution for particular positions across the ribbon. 
Consequences of these phenomena will appear in transport measurements due to the existence of four new conducting channels in the deformed area. Fig.\ \ref{Fig3} (c) compares conductance results, obtained with Green's function methods \cite{Leandro2018}, for ribbons with strains $0, 0.9\%$ and $1.7\%$. As expected, deformed ribbons (orange and green online) exhibit new conductance plateaus at energies corresponding to the  van Hove singularities that emerge in regions of increasing pseudo fields. The space separation between these channels is determined by the spread of the pseudo field that can be designed by choosing $b$, making it possible to collect selectively each valley current. These extra conducting channels are robust against edge disorder as the deformation resides inside the sample, away from disorder sources usually found at the edges of samples.

In conclusion, deformed graphene in the QH regime provides a perfect playground to create new valley polarized conducting channels. These appear whenever the sample is set up in the regime $l_B/b \ll 1$, at energies within LL gaps and at chosen locations in the sample. The separation of valley polarized states give rise to a unique braid pattern that should be observable in STM measurements of LDOS as the deformation is crossed. Hence, extended deformed graphene configurations offer novel and versatile setups to design electronic devices.

\noindent{\it Acknowledgments } We acknowledge discussions with Y. Jang, J. Mao,  E. Y. Andrei, and D. Zhai. This work was supported by IRTA-APS (DF, NS), CAPES (PrInt), CNPq  (AL, CL), FAPERJ E-26/202.953/2016 (AL) and E-26/202.768/2016 (LRFL), INCT de Nanomateriais de Carbono (AL, DF), NSF-DMR 1508325 (DF, NS). This work was partially performed at the Aspen Center for Physics, which is supported by NSF PHY-1607611.  Part of the calculations were done using the Ohio Supercomputer Center under the project PHS0265. DF is the Glidden Visiting Professor at OU.

\end{document}